\documentclass[preprint]{elsarticle}

\usepackage{lineno,hyperref}
\usepackage{extarrows}
\usepackage{amssymb}

\journal{J. Crystal Growth}

\begin{document}

\begin{frontmatter}

\date{January 12, 2018}

\title{On melt solutions for the growth of CaTiO$_3$ crystals}

\author[mymainaddress]{Detlef Klimm\corref{mycorrespondingauthor}}
\ead{detlef.klimm@ikz-berlin.de}

\author[mymainaddress,mysecondaryaddress]{Max Schmidt\fnref{fn1}}
\fntext[fn1]{This work is based on a thesis submitted by MS to the Department of Chemistry, Humboldt Universit\"at zu Berlin in partial fulfillment of the requirements for the degree of Master of Science.}

\author[mymainaddress]{Nora Wolff}

\author[mymainaddress]{Christo Guguschev}

\author[mymainaddress]{Steffen Ganschow}

\cortext[mycorrespondingauthor]{Corresponding author}

\address[mymainaddress]{Leibniz-Institut f\"ur Kristallz\"uchtung, Max-Born-Str. 2, 12489 Berlin, Germany}
\address[mysecondaryaddress]{Humboldt-Universit\"at zu Berlin, Institut f\"ur Chemie, Brook-Taylor-Str. 2, 12489 Berlin}

\begin{abstract}
When calcium titanate crystals are grown from stoichiometric melts, they crystallize in the cubic perovskite structure. Upon cooling to room temperature they undergo subsequent phase transitions to tetragonal and orthorhombic modifications. These phase transitions are disruptive and result in severely damaged crystals. This paper presents differential thermal analysis data for several prospective solvents, with the aim to identify a system offering the possibility to perform crystal growth of undistorted CaTiO$_3$ crystals by crystallizing them significantly below the melting point directly in the low temperature modification. From mixtures CaF$_2$:TiO$_2$:CaTiO$_3$ = 3:1:1 (molar ratio) the growth of undistorted, at least millimeter-sized CaTiO$_3$ crystals is possible.
\end{abstract}


\end{frontmatter}



\section{Introduction}

The chemical substance calcium titanate CaTiO$_3$ was reported first in 1840 by Rose as the mineral perovskite \cite{Rose40} and is now the prototype of the perovskite crystal structure. Although the mineral was first reported to belong to the regular (cubic) crystal system, this assumption is wrong: at least at room temperature a $Pbnm$ structure is stable. Instead, the cubic space group $Pm\bar{3}m$ is found for CaTiO$_3$ only at high temperatures. In this phase, Ti$^{4+}$ is octahedrally coordinated by six oxygen atoms and the larger Ca$^{2+}$ is placed in the center of a regular cuboctahedron of 12 oxygen atoms. Upon cooling the TiO$_6$ octahedra become distorted and rotated which leads in a series
\begin{equation} 
Pm\bar{3}m \xlongleftrightarrow{1523-1634\,\mathrm{K}} I4/mcm \xlongleftrightarrow{1423-\approx1523\,\mathrm{K}} Pbnm    \label{eq:PTs}
\end{equation}
to crystal structures with lower symmetry. The transformation temperatures in equation (\ref{eq:PTs}) are reported controversially in the literature \cite{Redfern96,Ali05,Yashima09}. Kennedy et al. \cite{Kennedy99} reported an additional $Cmcm$ phase between $Pbnm$ and $I4/mcm$ that could not be confirmed by other authors.

According to different authors \cite{Kaufman88,Jongejan70,DeCapitani98,FactSage7_1}, the $Pm\bar{3}m$ phase of CaTiO$_3$ melts congruently around $T_\mathrm{f}\approx2174-2262$\,K, but despite some reports \cite{Merker62} the phase transitions (\ref{eq:PTs}) result in twinned crystals during melt growth \cite{Jiang98}. The transition from the cubic to the tetragonal phase is lower for congruently melting Ba$_{1-x}$Ca$_x$TiO$_3$ ($x=0.227$) mixed crystals, but also there twinning occurs \cite{Kuper97}. It should be noted that CaTiO$_3$ is the only intermediate compound in the CaO--TiO$_2$ system that shows congruent melting. Two other intermediate compounds, both containing more CaO, undergo peritectic decomposition: Ca$_3$Ti$_2$O$_7$ melts peritectically under the formation of Ca$_4$Ti$_3$O$_{10}$ which later melts peritectically under the formation of CaTiO$_3$ \cite{Kaufman88,DeCapitani98}. These CaO-rich calcium titanates belong to a series of Ruddlesden-Popper phases \cite{Elcombe91}.

In cases where crystals cannot be grown from their congruent melts, often crystallization from high temperature melt solutions provides an alternative approach \cite{Elwell75}. The growth of millimeter-sized CaTiO$_3$ crystals was reported from different melt solutions based on ingredients like KF, B$_2$O$_3$, PbF$_2$, CaCl$_2\cdot2$\,H$_2$O, BaCl$_2\cdot2$\,H$_2$O, CaF$_2$, CsF, MoO$_3$ \cite{Watts89}. However it is a drawback of all these solvents that they add foreign ions to the melts that may enter up to a few percents in the crystal structure of the solute CaTiO$_3$. In recent studies the growth of high quality SrTiO$_3$ proved possible from an excess of TiO$_2$ as solvent \cite{Guguschev14,Guguschev17b}. The eutectic temperature of SrTiO$_3$/TiO$_2$ $T_\mathrm{eut}=1722\pm3$\,K is significantly lower than the congruent melting point of SrTiO$_3$ $T_\mathrm{f}=2352\pm20$\,K. Corresponding literature data for the CaTiO$_3$/TiO$_2$ eutectic range from $T_\mathrm{eut}=1690$\,K \cite{Kaufman88} to $T_\mathrm{eut}=1726$\,K \cite{DeCapitani98}, which indicates a less significant lowering of the liquidus temperature at the eutectic composition.

Good solvents are compounds that form a eutectic system with the solute, with vanishing solubility of the solvent in the solid phase of the solute \cite{Elwell75}. Besides this, the requirements for a good solvent are somewhat contradictory: On the one side, a low melting solvent results in a low $T_\mathrm{eut}$ and offers the possibility to perform the growth process at low $T$ -- possibly below the critical phase transitions (\ref{eq:PTs}). On the other side, the eutectic composition $x_\mathrm{eut}$ with a low melting solvent shifts close to the solvent, which lowers the yield of the growth process \cite{Klimm14c}. In this paper several solvents for the melt solution growth of CaTiO$_3$ crystals are tested on a thermoanalytic basis. The emphasis was on finding a solvent that allows a growth temperature below critical phase transformations, but with acceptable solubility for CaTiO$_3$. At the same time, the incorporation of foreign solvent ions should be avoided to the highest possible degree.


\section{Experimental}

Differential thermal analysis (DTA) with simultaneous thermogravimetry (TG) was performed using a NETZSCH STA 449C ``Jupiter'' thermal analyzer. A DTA/TG sample holder with Pt/Pt90Rh10 thermocouples and platinum crucibles allowed measurements up to 1920\,K in a flowing mixture of 20\,ml/min Ar + 20\,ml/min O$_2$. Only for the re-determination of the poorly defined melting point of pure CaTiO$_3$ (see previous section), a powder sample was melted in a NETZSCH STA 429C analyzer with W/Re thermocouples and tungsten crucibles. Usually the DTA samples were melted twice to ensure good mixing, and the second heating curves were used for further analysis.

Crystal phase and lattice parameter analysis of starting materials and of annealed samples was performed by X-ray diffraction using an XRD 3003 TT (GE Inspection Technologies). Cu K$_{\alpha1}$ radiation and a Bragg-Brentano setup were used with a scintillation detector. A heating stage by MRI (Materials Research Instruments) allowed XRD analysis up to 1600\,K.

The chemical composition of the grown crystals was investigated by micro X-ray fluorescence ($\mu$-XRF) measurements. The measurements were performed at low vacuum conditions (1 to 4\,mbar) using a Bruker M4 TORNADO spectrometer. The measurement system was equipped with a Rh X-ray source operated at 50\,kV and 200\,$\mu$A. Polycapillary X-ray optics were used to focus the Bremsstrahlung at the surface of the sample, which enabled a high spatial resolution of 25\,$\mu$m. The measurement time per point was set to 10\,s. The signals were detected using a circular silicon drift detector. For the accurate determination of the chemical composition, the XRF spectrometer was calibrated using a sintered CaTiO$_3$ pellet.


\section{Results and Discussion}

\subsection{Pure CaTiO$_3$}
\label{sec:pure}

\begin{figure}[ht]
\centering
\includegraphics[width=0.70\textwidth]{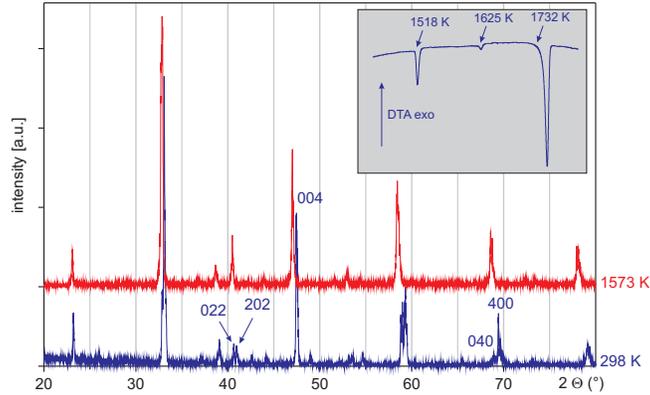}
\caption{X-ray powder patterns of CaTiO$_3$ in the orthorhombic phase at room temperature (298\,K) and in the tetragonal phase at 1573\,K. Orthorhombic indexing for space group setting $Pbnm$ \cite{Sasaki87}. Insert: DTA curve (5\,K/min, $2^\mathrm{nd}$ heating) of polycrystalline CaTiO$_3$.}
\label{fig:X-ray}
\end{figure}

The synthesis of CaTiO$_3$ powder was performed from a stoichiometric CaCO$_3$ (Fox Chemicals, 99.99\%)/TiO$_2$ (Alfa Aesar, 99.995\%) = 1:1 mixture. When this mixture is heated at 10\,K/min in the thermal analyzer, mass loss and endo\-thermal signal between $\approx1050$\,K and $\approx1200$\,K indicate the decomposition of CaCO$_3$ to CaO and CO$_2$. A subsequent sharp exothermal effect close to 1740\,K results from the reaction
\begin{equation}
\mathrm{CaO + TiO}_2 ~ \longrightarrow ~ \mathrm{CaTiO}_3 \qquad (\Delta H=-75.1\,\mathrm{kJ/mol\;@\;1740\,K})  \label{eq:formation}
\end{equation}
which proceeds without mass change. If the polycrystalline CaTiO$_3$ powder is heated repeatedly, small but reproducible endothermal peaks ($\lesssim300$\,J/mol, insert of Fig.~\ref{fig:X-ray}) around 1520\,K and 1625\,K indicate the orthorhombic/tetragonal and tetragonal/cubic phase transitions of the material (\ref{eq:PTs}). It should be noted that the latent heat for the first transition is significantly lower than reported in the FactSage database \cite{FactSage7_1}, but both values are very small compared to the heat of fusion (see below), with consequently large experimental error. The third DTA peak in Fig.~\ref{fig:X-ray} near 1723\,K coincides with the eutectic temperature between CaTiO$_3$ and TiO$_2$. One can assume that remaining inhomogeneity of the ceramic samples is responsible for this peak. The position of these DTA peaks reveals that simply working in a CaTiO$_3$--TiO$_2$ eutectic system cannot suppress the liquidus temperature below the phase transition temperatures.

A series of X-ray diffraction studies between room temperature and 1573\,K demonstrated the shift of X-ray peaks to smaller diffraction angles $2\,\Theta$, resulting from thermal expansion (Fig.~\ref{fig:X-ray}). Some peaks that are separated in the orthorhombic phase merge in the tetragonal phase, resulting from the higher symmetry (e.g. 022 and 202).

The melting behavior of this CaTiO$_3$ powder was investigated in the STA 429C and was compared for calibration to the melting of pure Al$_2$O$_3$ ($T_\mathrm{f}=2327$\,K, heat of fusion $\Delta H_\mathrm{f}=118.4$\,kJ/mol \cite{FactSage7_1}). From this measurement $T_\mathrm{f}=(2220\pm20)$\,K and $\Delta H_\mathrm{f}=113.3$\,kJ/mol were found for CaTiO$_3$. Even if a DTA measurement at such high temperatures gives usually only a rough estimation of $\Delta H_\mathrm{f}$ with errors of typically 20\%, this experimental value is in good agreement with FactSage \cite{FactSage7_1}, where 106.6\,kJ/mol are given.

\subsection{System CaTiO$_3$--KF}
\label{sec:KF}

Potassium fluoride was already successfully used for the melt solution growth of CaTiO$_3$ but has at least the drawback of high volatility which impedes long growth runs \cite{Watts89}. Besides, the molar solvent:solute ratio was 12:1, which limits significantly the yield of the crystallization process. Thus Watts et al. obtained from crucibles with 50\,ml volume starting at 1443\,K after 28 days only crystal sizes up to $2\times2\times2$\,mm$^3$ \cite{Watts89}.

\begin{figure}[ht]
\centering
\includegraphics[width=0.98\textwidth]{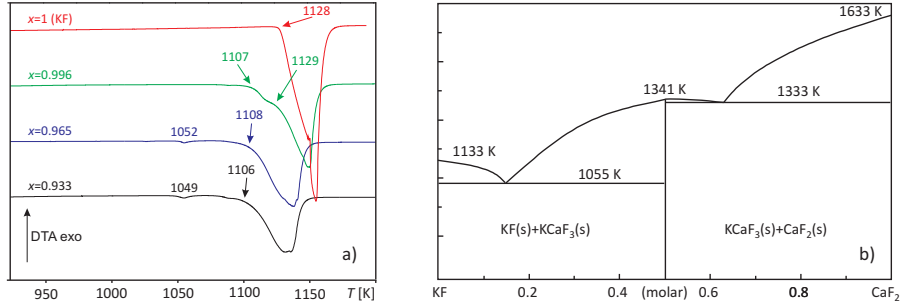}
\caption{a) DTA curves for the $2^\mathrm{nd}$ heating of different compositions starting from pure KF with $x=n\mathrm{[KF]}/(n\mathrm{[KF]}+n\mathrm{[CaTiO_3]})=0$. Onsets of DTA peaks are given in Kelvin. b) Phase diagram CaF$_2$--KF as published by Ishaque \cite{Ishaque52}.}
\label{fig:KF}
\end{figure}

To understand this growth process, DTA measurements of CaTiO$_3$--KF mixtures were performed up to ca. 1220\,K. This lower $T$ limit was chosen to avoid serious evaporation of KF that would otherwise shift the composition of DTA samples. Heating curves of four KF-rich compositions are given in Fig.~\ref{fig:KF}a) and show initially the expected drop of the liquidus temperature from pure KF, that hints at a CaTiO$_3$--KF eutectic. With higher concentration of the solute, starting with $x=0.965$, an unexpected small peak appears near 1050\,K. The lowermost curve in Fig.~\ref{fig:KF}a) almost corresponds to the melt growth solution that was used by Watts et al. \cite{Watts89}, except for a small CaO excess that was used there.

One can suspect that an disadvantageous ion exchange reaction of the kind
\begin{equation}
\mathrm{CaTiO}_3 \, + \,2\,\mathrm{KF} \rightleftarrows  \mathrm{CaF}_2 \, + \, \mathrm{K}_2\mathrm{TiO}_3  \label{eq:K2TiO3}
\end{equation}
produces some CaF$_2$ in the melt. Calcium fluoride is known to form with excess KF the perovskite type double fluoride KCaF$_3$ which is the only intermediate phase in the CaF$_2$--KF system \cite{Ishaque52,Hidaka84}. Because the publication of this system is rather old, several CaF$_2$/KF mixtures around the eutectic compositions in Fig.~\ref{fig:KF}b) were re-investigated by DTA. For the left eutectic, $(1054\pm2)$\,K, and for the right eutectic $(1328\pm3)$\,K were found which is a good confirmation of the literature data \cite{Ishaque52}.

The formation of CaF$_2$ (\ref{eq:K2TiO3}) and subsequently KCaF$_3$ reduces further the already small CaTiO$_3$ assay in the melt solution, which finally limits the efficiency of the growth process. Similar disadvantageous results were obtained with a series of other typical melt solution solvents, such as BaCl$_2$, SrF$_2$, NaF, CaCl$_2$, MgF$_2$. Details on these experiments can be found in the accompanying material.

\subsection{System CaTiO$_3$--CaF$_2$--TiO$_2$}
\label{sec:ternary}

The results of the previous section~\ref{sec:KF} showed that the application of solvents based on foreign cations (such as K$^+$) is critical because they might give rise to parasitic phases. Besides the formation of solid solutions between solvent and solute is a typical issue that was observed e.g. for growth from solutions containing PbF$_2$ where Pb$^{2+}$ substitutes partially for Ca$^{2+}$ in CaTiO$_3$ \cite{Watts89}.

\begin{figure}[ht]
\centering
\includegraphics[width=0.98\textwidth]{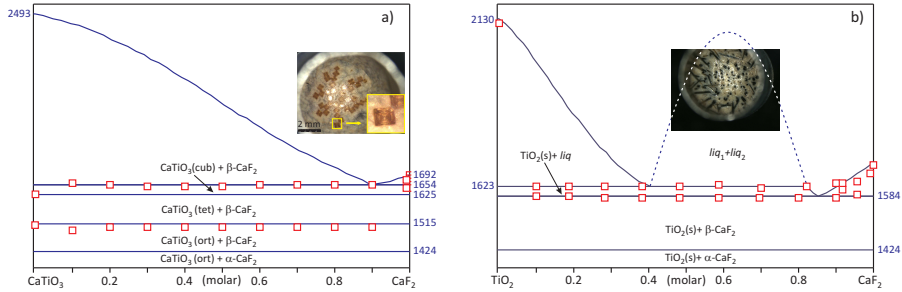}
\caption{a) Experimental points and thermodynamic assessment \cite{FactSage7_1} of the CaF$_2$--CaTiO$_3$ system. The insert shows the DTA crucible after measuring an almost eutectic ($x\approx0.9$) sample with CaTiO$_3$ crystallites. b) Experimental points and thermodynamic assessment of the CaF$_2$--TiO$_2$ system. The insert shows the DTA crucible after measuring a sample $x\approx0.6$ with TiO$_2$ (rutile) needles.}
\label{fig:CaF2}
\end{figure}

In a series of DTA measurements the pseudobinary system CaTiO$_3$--CaF$_2$ was investigated up to 1723\,K and it was found to be eutectic. Fig.~\ref{fig:CaF2}a) shows that the liquidus drops from the melting point of pure CaF$_2$ ($T_\mathrm{f}=(1692\pm3)$\,K) to $T_\mathrm{eut}=(1654\pm3)$\,K which is reached at $x_\mathrm{eut}\approx0.9$ CaF$_2$. The isotherm at 1424\,K marks the $\alpha/\beta$ transition of CaF$_2$ which has, however, only a marginal thermal effect \cite{Mirwald78}. $T_\mathrm{eut}$ is 29\,K higher than the highest \textit{t/c} transition of CaTiO$_3$. For samples $x\lesssim x_\mathrm{eut}$, the liquidus temperature can be exceeded during the DTA measurements and during cooling rectangular crystallites with dimensions $\leq0.5$\,mm are formed. Their surfaces, however, are concave stepped and microscopic inspection shows typical flaws resulting from phase transitions that occurred after crystallization.

It was mentioned in the introduction that for SrTiO$_3$ the growth temperature could be lowered substantially using melt compositions that are close to the SrTiO$_3$/TiO$_2$ eutectic. Because the corresponding reduction of the liquidus temperature is insufficient for the growth of CaTiO$_3$, it seemed worthwhile to study a combination of CaF$_2$ and TiO$_2$ as prospective solvent. In a first step, the pseudobinary system CaF$_2$--TiO$_2$ had to be investigated.

Hillert \cite{Hillert65} published a CaF$_2$--TiO$_2$ phase diagram based on annealing experiments and found it to be eutectic with a monotectic miscibility gap in the liquid. In his study $x_\mathrm{eut}$ is almost in the middle of the system, on the TiO$_2$-rich side of the monotectic. Fig.~\ref{fig:CaF2}b) shows the present DTA results that are contradictory: The monotectic line at $T=1623$\,K extends to the TiO$_2$ side of the system and $x_\mathrm{eut}$ is CaF$_2$-rich. Besides, both isothermal lines are significantly lower in the present study: $T_\mathrm{eut}=(1584\pm2)$\,K rather than 1633\,K \cite{Hillert65}, and monotectic demixing at $(1623\pm3)$\,K rather than 1638\,K \cite{Hillert65}.

One can expect as a result of the opposite position of $x_\mathrm{eut}$ that the stable form of solid TiO$_2$ (tetragonal rutile) has a wide range of primary crystallization in the phase diagram. Indeed, visual inspection of DTA samples with $0.4\leq x\leq0.7$ showed dark rutile needles at the melt surface. One example is shown in the insert of Fig.~\ref{fig:CaF2}b).

\begin{figure}[ht]
\centering
\includegraphics[width=0.60\textwidth]{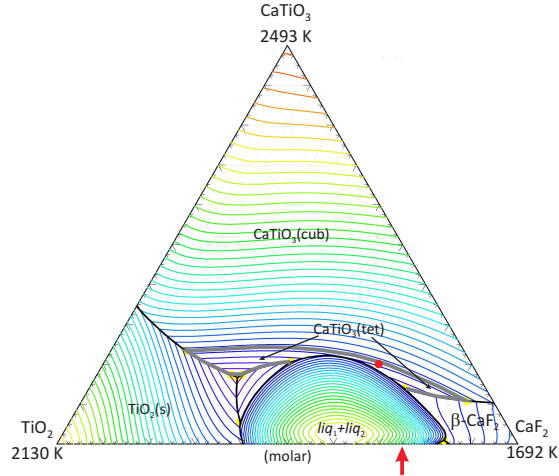}
\caption{Liquidus projection on the CaTiO$_3$--CaF$_2$--TiO$_2$ ternary system with 20\,K difference between isotherms. The red arrow at the bottom marks the solvent composition $x'=0.75$\,CaF$_2$ + 0.25\,TiO$_2$.}
\label{fig:ternary}
\end{figure}

The eutectic temperature in Fig.~\ref{fig:CaF2}b) is 41\,K below the tetragonal/cubic transition of CaTiO$_3$ that was measured in section~\ref{sec:pure}. This raises the hope that a mixture of CaF$_2$ and TiO$_2$ could be used as a solvent from which CaTiO$_3$ can be crystallized at least below that transition. In a series of DTA measurements, to samples starting from ``pure solvent'' 0.75\,CaF$_2$ + 0.25\,TiO$_2$ (set to $x'=1$), increasing amounts of CaTiO$_3$ were added up to
\[x'=\frac{n(0.75\,\mathrm{CaF}_2+0.25\,\mathrm{TiO}_2)}{n{(0.75\,\mathrm{CaF}_2+0.25\,\mathrm{TiO}_2)+n(\mathrm{CaTiO}_3)}}=0.79
\]
($n$ -- corresponding molar quantities). The experimental results of this study, together with published data on the CaO--TiO$_2$ binary, allowed the construction of the concentration triangle CaTiO$_3$--CaF$_2$--TiO$_2$ that is shown in Fig.~\ref{fig:ternary}.

To calculate this graph, data for the solids CaTiO$_3$ (cub, tet, ort), CaF$_2$ ($\alpha$-, $\beta$-), and TiO$_2$ were taken from the FactSage \cite{FactSage7_1} databases if possible. A sufficient model for the excess Gibbs free energy of the melt was found by Redlich-Kister \cite{Redlich48} polynomials for the binaries
\begin{equation}
G^\mathrm{ex} = \sum_{i=0}^n ~^i\!L_\mathrm{AB} \, x_\mathrm{A} \, x_\mathrm{B} (x_\mathrm{A}-x_\mathrm{B})^i      \label{eq:G_ex}
\end{equation}
with parameters $^i\!L_\mathrm{AB}$ that are given in Table~\ref{tab:Parameter}. It should be noted that these parameters were not the subject of a detailed numerical assessment, because insufficient experimental data were available, especially inside the ternary region. Nevertheless a reasonable description of all DTA results was possible on this basis even without using additional ternary interaction parameters.

\begin{table}[ht]
\centering
\begin{tabular}{lrrrr}
System                  & $i=0$    & $i=1$    & $i=2$    & $i=8$    \\
\hline
$L$(CaTiO$_3$--CaF$_2$) &  -3000   &  -10000  &  10000   &  8000   \\
$L$(CaF$_2$--TiO$_2$)   &  11500   &  20500   &  -16000  & --       \\
$L$(CaTiO$_3$--TiO$_2$) &  -19500  &  --      & --       & -- \\
\hline
\end{tabular}
\caption{Redlich-Kister \cite{Redlich48} parameters $^i\!L_\mathrm{AB}$ for binary interactions in the liquid phase (\ref{eq:G_ex}).}
\label{tab:Parameter}
\end{table}

As expected, the cubic modification of CaTiO$_3$ crystallizes from melt compositions that are close to this corner of the concentration triangle Fig.~\ref{fig:ternary}. The insert of Fig.~\ref{fig:CaF2}a) showed already that primary crystallization of this phase results in distorted crystallites because they undergo a critical phase transition upon cooling. Below the bold gray isotherm, however, CaTiO$_3$(tet) crystallizes first, and the insert of Fig.~\ref{fig:CaF2}b) showed that such crystallites are less prone to distortions. This primary crystallization field of CaTiO$_3$(tet) is partially covered by a phase field where demixing of the melt occurs. Fig.~\ref{fig:CaF2}b) shows that also there the system is completely molten for $T>1623$\,K which is sufficient to prevent crystallization of CaTiO$_3$(cub). Indeed, the DTA samples with compositions inside the primary crystallization field of CaTiO$_3$(tet) contained small ($\leq0.5$\,mm) CaTiO$_3$ crystallites that were free of distortions.

\begin{figure}[ht]
\centering
\includegraphics[width=0.40\textwidth]{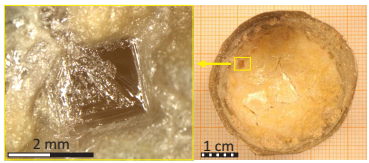}
\caption{First CaTiO$_3$ crystals obtained by unseeded growth in a 30\,ml crucible. The starting composition is marked by a red dot in Fig.~\ref{fig:ternary}.}
\label{fig:X-tal}
\end{figure}

In a first unseeded crystal growth experiment similar conditions were checked on a larger scale. A 30\,ml Pt crucible (diameter 40\,mm, height 35\,mm) was filled with 15.62\,g of a 3:1:1 (molar) mixture of CaF$_2$, TiO$_2$ and CaTiO$_3$ powders, and was covered by a Pt lid. In a muffle furnace the crucible was heated at 5\,K/min to 1673\,K and held there 3\,h for homogenization. Then the crucible was cooled at a rate of 4\,K/h to 1073\,K where the heating power was switched off.

Fig.~\ref{fig:X-tal} shows a photograph of the crucible after this process, together with the magnification of one of the crystals that could be found in the solidified melt. These crystals are located in the vicinity of the crucible wall. Most of them are cubes or cuboids with edge length up to 2.5\,mm; all of them are free of the flaws that were found in crystals grown at higher temperature. Even if the crystals are clear, they exhibit a bright brown coloring. X-ray fluorescence analysis revealed a stoichiometric composition since, within the range of typical measurement uncertainties, no signiﬁcant differences from the sintered standard pellet were detected. It turned out, unfortunately, that it is very difficult to remove crystals from the melt body.


\section{Discussion and Conclusions}

It could be shown that a combination of TiO$_2$ and CaF$_2$ represents a suitable solvent for solution growth of CaTiO$_3$ crystals. This combination has the benefit that foreign cations are completely omitted, and F$^-$ is the only foreign ion at all. A molar ratio solute (CaTiO$_3$) to solvent (CaF$_2$+TiO$_2$) of 1:4 can be used for the growth process which is considerably better than for other solvents such as KF (see section~\ref{sec:KF}) were the ratio is in the order of 1:12.

Melts with high concentrations of CaF$_2$ and TiO$_2$ exhibit a monotectic miscibility gap, which is shown near the bottom line of the concentration triangle in Fig.~\ref{fig:ternary}, or for the CaF$_2$--TiO$_2$ system without CaTiO$_3$ in Fig.~\ref{fig:CaF2}b, respectively. The corresponding ``\textit{liq$_1$+liq$_2$}'' phase fields are beneficial for crystal growth from melt solution, because they reduce significantly the temperature where the first precipitation of a solid phase from the melt(s) occurs. In Fig.~\ref{fig:CaF2}b already for $x\geq0.4$ the whole system is liquid above 1623\,K. This would not be the case if the system were simply eutectic, because the melting point of TiO$_2$ is rather high.

\begin{figure}[ht]
\centering
\includegraphics[width=0.50\textwidth]{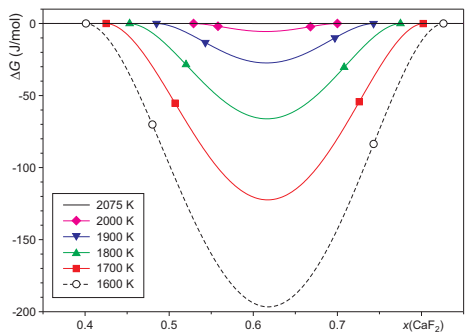}
\caption{Gibbs energy reduction $\Delta G=G^\mathrm{demixed}-G^\mathrm{mixed}$ of melts with composition $x$\,CaF$_2+(1-x)$\,TiO$_2$ in the vicinity of the ``$liq_1+liq_2$'' phase field of Fig.~\ref{fig:CaF2}b) for different temperatures.}
\label{fig:demixing}
\end{figure}

A positive (repulsive) excess Gibbs free energy contribution $\Delta G$ which can be described by the parameters in Table~\ref{tab:Parameter} is the origin of the miscibility gap. The total amount of this ``demixing'' contribution to $G^\mathrm{ex}$ is shown in Fig.~\ref{fig:demixing} and exceeds never 200\,J/mol which is so small, that an experimental detection by normal DTA measurements at high $T\geq1600$\,K is difficult. According to the experimental data that are presented here, the positive $\Delta G$ shrinks as to expected with $T$ and disappears at $T\approx2075$\,K. The inflection points which are marked by symbols in the $\Delta G(x)$ functions in Fig.~\ref{fig:demixing} mark the points where spinodal decomposition of the melt occurs. The outer symbols on the curves mark the equilibrium concentration where demixing begins at the corresponding temperature.

From the thermodynamic dataset that was used for the calculation of Fig.~\ref{fig:ternary} the ternary eutectic point between CaTiO$_3$, CaF$_2$ and TiO$_2$ is at $T_\mathrm{eut}=1522$\,K and just 8\,K below the tetragonal/orthorhombic phase transition of CaTiO$_3$ which comes out in this calculation at 1530\,K. The minor CaTiO$_3$(ortho) phase field is, however, not labeled in this figure because it is very small and hard to recognize. So far it cannot be decided if the growth process that led to the crystals shown in Fig.~\ref{fig:X-tal} were crystallized in the tetragonal phase and transformed to the room temperature orthorhombic phase without distortion, or if they crystallized directly in the stable orthorhombic phase. Possibly also crystallization inside the CaTiO$_3$(tet) phase field could lead to the immediate crystallization of CaTiO$_3$(ort), if supercooling occurs. Severe supercooling is known to be a common phenomenon during oxide crystal growth, including titanates \cite{Wanklyn84b,Nabokin03}.

\section*{Acknowledgements}

The authors are grateful to Albert Kwasniewski for performing X-ray measurements. We thank Detlev Schulz for reading the manuscript.

\section*{References}


\begin{thebibliography}{10}

\bibitem{Rose40}
G.~Rose, {\"Uber einige neue Mineralien des Urals}, {J. f\"ur Praktische Chemie} 19 (1840) 459--468.
\newblock \href {http://dx.doi.org/10.1002/prac.18400190179} {\path{doi:10.1002/prac.18400190179}}.

\bibitem{Redfern96}
S.~A.~T. Redfern, High-temperature structural phase transitions in perovskite {(CaTiO$_3$)}, Phase Transitions 8 (1996) 8267--8275.
\newblock \href {http://dx.doi.org/10.1088/0953-8984/8/43/019} {\path{doi:10.1088/0953-8984/8/43/019}}.

\bibitem{Ali05}
R.~Ali, M.~Yashima, Space group and crystal structure of the perovskite {CaTiO$_3$} from 296 to 1720\,{K}, J. Sol. State Chem. 178 (2005) 2867--2872.
\newblock \href {http://dx.doi.org/10.1016/j.jssc.2005.06.027} {\path{doi:10.1016/j.jssc.2005.06.027}}.

\bibitem{Yashima09}
M.~Yashima, R.~Ali, Structural phase transition and octahedral tilting in the calcium titanate perovskite {CaTiO$_3$}, Solid State Ionics 180 (2009) 120--126.
\newblock \href {http://dx.doi.org/10.1016/j.ssi.2008.11.019} {\path{doi:10.1016/j.ssi.2008.11.019}}.

\bibitem{Kennedy99}
B.~J. Kennedy, C.~J. Howard, B.~C. Chakoumakos, Phase transitions in perovskite at elevated temperatures -- a powder neutron diffraction study, J. Phys.: Condens. Matt. 11 (1999) 1479--1488.
\newblock \href {http://dx.doi.org/10.1088/0953-8984/11/6/012} {\path{doi:10.1088/0953-8984/11/6/012}}.

\bibitem{Kaufman88}
L.~Kaufman, Calculation of multicomponent ceramic phase diagrams, Physica B 150 (1988) 99--114.
\newblock \href {http://dx.doi.org/10.1016/0378-4363(88)90111-8} {\path{doi:10.1016/0378-4363(88)90111-8}}.

\bibitem{Jongejan70}
A.~Jongejan, A.~Wilkins, A re-examination of the system {CaO--TiO$_2$} at liquidus temperatures, J. Less Common Metals 20~(4) (1970) 273--279.
\newblock \href {http://dx.doi.org/https://doi.org/10.1016/0022-5088(70)90001-9} {\path{doi:https://doi.org/10.1016/0022-5088(70)90001-9}}.

\bibitem{DeCapitani98}
C.~DeCapitani, M.~Kirschen, A generalized multicomponent excess function with application to immiscible liquids in the system {CaO--SiO$_2$--TiO$_2$},
  Geochimica et Cosmochimica Acta 62 (1998) 3753--3763.
\newblock \href {http://dx.doi.org/10.1016/S0016-7037(98)00319-6} {\path{doi:10.1016/S0016-7037(98)00319-6}}.

\bibitem{FactSage7_1}
www.factsage.com, {FactSage 7.1}, GTT Technologies, Kaiserstr. 100, 52134 Herzogenrath, Germany (2017).

\bibitem{Merker62}
L.~Merker, Synthesis of calcium titanate single crystals by flame fusion technique, J. Amer. Ceram. Soc. 45 (1962) 366--369.
\newblock \href {http://dx.doi.org/10.1111/j.1151-2916.1962.tb11170.x} {\path{doi:10.1111/j.1151-2916.1962.tb11170.x}}.

\bibitem{Jiang98}
Y.~Jiang, R.~Guo, A.~S. Bhalla, Growth and properties of {CaTiO$_3$} single crystal fibers, J. Electroceram. 2 (1998) 199--203.
\newblock \href {http://dx.doi.org/10.1023/A:1009978901009} {\path{doi:10.1023/A:1009978901009}}.

\bibitem{Kuper97}
C.~Kuper, R.~Pankrath, H.~Hesse, Growth and dielectric properties of congruently melting {Ba$_{1-x}$Ca$_x$TiO$_3$} crystals, Appl. Phys. A: Mat. Sci. {\&} Processing 65 (1997) 301--305.
\newblock \href {http://dx.doi.org/10.1007/s003390050583} {\path{doi:10.1007/s003390050583}}.

\bibitem{Elcombe91}
M.~M. Elcombe, E.~H. Kisi, K.~D. Hawkins, T.~J. White, P.~Goodman, S.~Matheson, Structure determinations for {Ca$_3$Ti$_2$O$_7$, Ca$_4$Ti$_3$O$_{10}$,
  Ca$_{3.6}$Sr$_{0.4}$Ti$_3$O$_{10}$} and a refinement of {Sr$_3$Ti$_2$O$_7$}, Acta Cryst. B 47 (1991) 305--314.
\newblock \href {http://dx.doi.org/10.1107/S0108768190013416} {\path{doi:10.1107/S0108768190013416}}.

\bibitem{Elwell75}
D.~Elwell, H.~J. Scheel, Crystal Growth from High Temperature Solutions, Academic Press, London, 1975.
\newblock \href {http://dx.doi.org/10.3929/ethz-a-006779537} {\path{doi:10.3929/ethz-a-006779537}}.

\bibitem{Watts89}
B.~E. Watts, H.~Dabkowska, B.~M. Wanklyn, The flux growth of perovskites ({CaTiO$_3$, CdTiO$_3$, SrZrO$_3$}, and {LaGaO$_3$, PrGaO$_3$, NdGaO$_3$}), J. Cryst. Growth 94 (1989) 125 -- 130.
\newblock \href {http://dx.doi.org/10.1016/0022-0248(89)90611-8} {\path{doi:10.1016/0022-0248(89)90611-8}}.

\bibitem{Guguschev14}
C.~Guguschev, D.~Klimm, F.~Langhans, Z.~Galazka, D.~Kok, U.~Juda, R.~Uecker, Top-seeded solution growth of {SrTiO$_3$} crystals and phase diagram studies
  in the {SrO--TiO$_2$} system, CrystEngComm 16 (2014) 1735--1740.
\newblock \href {http://dx.doi.org/10.1039/c3ce42037j} {\path{doi:10.1039/c3ce42037j}}.

\bibitem{Guguschev17b}
C.~Guguschev, D.~J. Kok, U.~Juda, R.~Uecker, S.~Sintonen, Z.~Galazka, M.~Bickermann, Top-seeded solution growth of {SrTiO$_3$} single crystals virtually free of mosaicity, J. Crystal Growth 468 (2017) 305--310.
\newblock \href {http://dx.doi.org/10.1016/j.jcrysgro.2016.10.048} {\path{doi:10.1016/j.jcrysgro.2016.10.048}}.

\bibitem{Klimm14c}
D.~Klimm, {Phase Equilibria}, in: T.~Nishinaga (Ed.), Handbook of Crystal Growth, $2^\mathrm{nd}$ Edition, Elsevier, 2014, pp. 85--136.
\newblock \href {http://dx.doi.org/10.1016/B978-0-444-56369-9.00002-2} {\path{doi:10.1016/B978-0-444-56369-9.00002-2}}.

\bibitem{Sasaki87}
S.~Sasaki, C.~T. Prewitt, J.~D. Bass, W.~A. Schulze, Orthorhombic perovskite {CaTiO$_3$} and {CdTiO$_3$}: structure and space group, Acta Cryst. C 43 (1987) 1668--1674.
\newblock \href {http://dx.doi.org/10.1107/S0108270187090620} {\path{doi:10.1107/S0108270187090620}}.

\bibitem{Ishaque52}
M.~Ishaque, Equilibres liquide-solide dans le systeme quaternaire --- {ClNa ClK Cl$_2$Ca FNa FK F$_2$Ca} --- les 3 systemes ternaires reciproques correspondants, le systeme ternaire aux 3 fluorures, le systeme binaire {FK-F$_2$Ca}, {Bull. Soc. Chim. de France} 19 (1952) 127--138, cited from {``Phase Diagrams for Ceramists'', entry 3348}.

\bibitem{Hidaka84}
M.~Hidaka, S.~Yamashita, Y.~Okamoto, Study of structural phase transitions of {KCaF$_3$}, phys. stat. sol. (a) 81 (1984) 177--183.
\newblock \href {http://dx.doi.org/10.1002/pssa.2210810117} {\path{doi:10.1002/pssa.2210810117}}.

\bibitem{Mirwald78}
P.~W. Mirwald, G.~C. Kennedy, The phase relations of calcium fluoride (fluorite) to 60\,kbars and 1800$^{\,\circ}${C}, J. Phys. Chem. Sol. 39 (1978) 859--861.
\newblock \href {http://dx.doi.org/10.1016/0022-3697(78)90145-2} {\path{doi:10.1016/0022-3697(78)90145-2}}.

\bibitem{Hillert65}
L.~Hillert, The phase diagram {TiO$_2$--CaF$_2$}, Acta Chemica Scandinavica 19 (1965) 1516, see also {``Phase Diagrams for Ceramists'', entry 4879}.
\newblock \href {http://dx.doi.org/10.3891/acta.chem.scand.19-1516} {\path{doi:10.3891/acta.chem.scand.19-1516}}.

\bibitem{Redlich48}
O.~Redlich, A.~T. Kister, Algebraic representation of thermodynamic properties and the classification of solutions, Industrial {\&} Engineering Chemistry 40 (1948) 345--348.
\newblock \href {http://dx.doi.org/10.1021/ie50458a036} {\path{doi:10.1021/ie50458a036}}.

\bibitem{Wanklyn84b}
B.~M. Wanklyn, B.~E. Watts, Nucleation in flux growth of oxides: {DTA} and {TGA} experiments, J. Crystal Growth 70 (1984) 459--465.
\newblock \href {http://dx.doi.org/10.1016/0022-0248(84)90302-6} {\path{doi:10.1016/0022-0248(84)90302-6}}.

\bibitem{Nabokin03}
P.~I. Nabokin, D.~Souptel, A.~M. Balbashov, Floating zone growth of high-quality {SrTiO$_3$} single crystals, J. Crystal Growth 250 (2003) 397--404.
\newblock \href {http://dx.doi.org/10.1016/S0022-0248(02)02391-6} {\path{doi:10.1016/S0022-0248(02)02391-6}}.

\end{thebibliography}

\end{document}